\documentclass[runningheads]{llncs}

\usepackage[T1]{fontenc}
\newif\iffinal
\finalfalse

\usepackage[table]{xcolor}
\usepackage{calc}
\usepackage[framemethod=TikZ]{mdframed}
\usepackage{forest}
\usepackage{subcaption}
\usepackage{xspace}
\usepackage{courier}
\usepackage{makecell}
\usepackage{svg}
\usepackage{booktabs,makecell}
\usepackage[hidelinks]{hyperref}
\usepackage[altpo,epsilon]{backnaur}
\usepackage{amsmath} 
\usepackage[capitalise]{cleveref}
\usepackage{mathtools}
\usepackage{caption}
\usepackage{textcomp}  
\usepackage{multirow}
\usepackage{paralist}
\usepackage{balance}
\usepackage{dirtytalk}
\usepackage{afterpage}
\usepackage{adjustbox} 
\usepackage{backnaur}
\usepackage{tikz}
\usetikzlibrary{tikzmark,calc,shapes, positioning,arrows.meta}

\crefname{appendix}{App.}{Apps.}
\crefname{observation}{observation}{observations}
\Crefname{observation}{Observation}{Observations}

\usepackage{enumitem,amssymb} 
\newlist{todolist}{itemize}{2}
\setlist[todolist]{label=$\square$}
\usepackage[l3]{csvsimple}  
\usepackage{listings,array,varwidth,listings-rust}
\usepackage{lstautogobble}  
\usepackage{color}          
\usepackage{zi4}            
\definecolor{darkgreen}{RGB}{0, 150, 0}
\definecolor{bluekeywords}{rgb}{0.13, 0.13, 1}
\definecolor{greencomments}{rgb}{0, 0.5, 0}
\definecolor{redstrings}{rgb}{0.9, 0, 0}
\definecolor{graynumbers}{rgb}{0.5, 0.5, 0.5}
\definecolor{pastelpink}{RGB}{255, 182, 193}
\definecolor{pastelblue}{RGB}{173, 216, 230}
\definecolor{pastelgreen}{RGB}{182, 255, 182}
\definecolor{pastelyellow}{RGB}{255, 255, 204}
\definecolor{pastelpurple}{RGB}{216, 191, 216}
\definecolor{pastelmustard}{RGB}{255, 219, 88}
\usepackage{listings}
\newcommand{\bmc}{\text{BMC}\xspace}

\newcommand{\seahorn}{\textsc{SeaHorn}\xspace}

\newcommand{\klee}{\textsc{KLEE}\xspace}

\newcommand{\llvm}{\textsc{LLVM}\xspace}

\newcommand{\smack}{\textsc{SMACK}\xspace}
\newcommand{\crux}{\text{Crux}\xspace}

\newcommand{\smt}{\textsc{SMT}\xspace}

\newcommand{\sut}{\textsc{SUT}\xspace}
\newcommand{\env}{\text{environment}\xspace}

\newcommand{\fak}{\text{fake}\xspace}

\newcommand{\otoh}{\text{On the other hand}\xspace}

\newcommand{\mbedtls}{\text{mbedTLS}\xspace}
\newcommand{\rust}{\text{Rust}\xspace}

\newcommand{\seamock}{\textsc{SeaMock}\xspace}

\newlength{\markoffset}
\iffinal
	\newcommand{\ag}[1]{\textcolor{purple}{}\xspace}
	\newcommand{\sid}[1]{\textcolor{darkgreen}{}\xspace}

\else
	\newcommand{\ag}[1]{\textcolor{purple}{[AG: #1]}\xspace}
	\newcommand{\sid}[1]{\textcolor{darkgreen}{[SP: #1]}\xspace}
\fi

\newcommand{\code}[1]{\lstinline[basicstyle=\fontsize{8pt}{9pt}\ttfamily,style=SeaC++] {#1}\xspace}

\newcommand{\bparagraph}[1]{\noindent\textbf{#1}}
\newcommand{\lref}[1]{line \ref{#1}}
\newcommand{\Lref}[1]{Line \ref{#1}}

\newcommand{\totalfuncs}{{44}\xspace}
\newcommand{\verifiedfuncs}{{31}\xspace}
\newcommand{\notverifiedfuncs}{{13}\xspace}
\newcommand{\trivialfuncs}{{9}\xspace}
\newcommand{\notinterestedfuncs}{{4}\xspace}

\newcommand\sdef{\ensuremath{\triangleq}}

\lstdefinelanguage{seair}%
{morekeywords={abstract,%
			assume,sassert,assert,phi,br,%
			alloca,malloc,load,store,%
			gep,halt,select,%
			isderef,isalloc,ismod,resetmod%
			fun,main,add,ptoi,itop,free,memmove,memcpy%
		},%
	numbers=none, 
	mathescape=true,%
	sensitive,%
	commentstyle=\color{blue},
	morecomment=[l]//,%
	morecomment=[s]{/*}{*/},%
	morestring=[b]",%
	morestring=[b]',%
	showstringspaces=false
}[keywords,comments,strings]%

\lstdefinestyle{SeaC++}{
    language=C++,
    morekeywords={size_t,assume, sassert},
}
\lstdefinestyle{SeaC}{
    language=C,
    morekeywords={size_t,assume, sassert},
}

\newcommand{\cvtype}{\ensuremath{C_{v}}\xspace}
\newcommand{\cstype}{\ensuremath{C_{s}}\xspace}
\newcommand{\cftype}{\ensuremath{C_{f}}\xspace}
\newcommand{\cntype}{\ensuremath{C_{n}}\xspace}

\spnewtheorem{observation}{Observation}{\bfseries}{\itshape}

\newif\ifarxivmode
\arxivmodetrue

\ifarxivmode
\else
\usepackage[firstpage]{draftwatermark}
\SetWatermarkText{\hspace*{340bp}\raisebox{620bp}{\href{}{\includegraphics[scale=0.3]{ArtifactsAvailable}}\hspace{2bp}\href{https://sefm-conference.github.io/2024/artifacts/}{\includegraphics[scale=0.3]{ArtifactsEvaluatedFunctional}}}}
\SetWatermarkAngle{0}
\fi
\begin{document}
\title{Unlocking the Power of Environment Assumptions for Unit Proofs}
\author{%
	Siddharth Priya\inst{1}
	\and Temesghen Kahsai\inst{2}
	\and Arie Gurfinkel\inst{1}
}
\authorrunning{S. Priya et al.}

\institute{University of Waterloo\inst{1} and Amazon\inst{2}}

\maketitle

\begin{abstract}
	Clearly articulating the assumptions of the execution environment is crucial for the successful application of code-level formal verification. The process of specifying a model for the environment can be both laborious and error-prone, often requiring domain experts. In contrast, when engineers write unit tests, they frequently employ mocks (tMocks) to define the expected behavior of the environment in which the function under test operates. These tMocks describe how the environment behaves, e.g., the return types of an external API call (stateless behaviour) or the correct sequence of function calls (stateful behaviour). Mocking frameworks have proven to be highly effective tools for crafting unit tests. In our work, we draw inspiration from tMocks and introduce their counterpart in the realm of formal verification, which we term \emph{vMocks}. vMocks offer an intuitive framework for specifying a \emph{plausible} environment when conducting code-level formal verification. We implement a vMock library for the verification of C programs called \seamock.
	We investigate the practicality of vMocks by, first, comparing specifications styles in the communication layer of the Android Trusty Trusted Execution Environment (TEE) open source project, and second, in the verification of \mbedtls, a widely used open source C library that provides secure communication protocols and cryptography primitives for embedded systems.
	Based on our experience, we conclude that vMocks complement other forms of environment models. We believe that vMocks ease adoption of code-level formal verification among developers already familiar with tMocks.
\end{abstract}

\section{Introduction}
\label{sec:vmocks-intro}

Formal verification can provide stronger guarantees for software than testing.
To scale and efficiently perform code-level formal verification, it is imperative to establish the boundary of a verification task. This delineation defines what falls within the purview of the verification and what lies beyond, essentially giving rise to two core components: the System Under Verification (\sut\footnote{We use the more familiar SUT as an acronym instead of the more accurate SUV.}) and the assumptions governing the SUT's operating environment.

Code-level formal verification in industry is typically lead by verification experts and takes months to become reliable.
Even with expertise, effective environment models can take months to years to fine-tune in the context of complex industrial codebases~\cite{DBLP:conf/fmcad/LahiriLGNLKDLB20,DBLP:conf/eurosys/BallBCLLMORU06}. The difficulty of designing environments is specially pronounced in new development projects where a reference implementation of the environment may be unavailable.

Contrarily, in test driven development (TDD), developers use testing mocking frameworks (a.k.a., tMocks) to implement environments for SUTs.
tMock environments do not specify a complete environment.
Instead, they are plausible only for a single function (the \sut) in a unit test.
This methodology circumvents the problem of defining a complete test environment operationally (which may be complex to get right and may change over time) by instead describing how it \emph{behaves} (e.g., how many times an environment function may be invoked or the sequence of environment function calls).

Inspired by the success of tMocks, we introduce vMocks -- a mocking framework for code-level formal verification. Developers can use vMocks to clearly and efficiently model the environment for the \sut when performing code-level verification. We believe that vMocks will be transformative in enabling the adoption of code-level formal verification early in the software development cycle. 

The practice of mocking rests on an effective Domain Specific Language (DSL) for describing externally visible \emph{actions} taken by a mock.
It is important to have the right DSL design and implementation. First, the DSL must be \emph{concise} since each unit test has its own environment and a verbose language may result in developer fatigue and introduce specification errors. Second, the DSL must be \emph{embedded} in the host programming language to avoid complexity of translating between the DSL semantics and the SUT. Finally, the specification (mock) must be \emph{executable} with the chosen testing framework.

Existing tMock frameworks satisfy these characteristic with their intended usage scenario.
For example, frameworks like cMocka~\cite{CMocka} and GoogleMock~\cite{GoogleTest} for C and C++ respectively offer concise DSLs for concrete execution, are embedded in the programming language, and are executable with unit testing frameworks.
However, they do not fit well with verification tools.
Such tools employ symbolic execution environments.
Here, a single unit proof\footnote{A \emph{unit proof} is a symbolic unit test used in code-level verification.} may explore multiple concrete executions and thus the DSL must express non-determinism in the mock environment.
Additionally, valid design choices for concrete execution contexts can become very expensive in symbolic executions.
For example, cMocka uses a runtime map of function names to mock actions that is queried on each function call to the mock environment.
GoogleMock uses dynamic dispatch to wire virtual calls to corresponding mock implementation.
Therefore, using existing frameworks has considerable friction.

To address these problems, we have developed, to our knowledge, the first, compile-time mocking framework, called \seamock.
Its GoogleMock-inspired DSL uses C++ meta-programming to provide actions (including non--determinism) to describe vMock behaviour.
However, unlike GoogleMock, all function calls are resolved at compile time.
Thus, no cost is incurred at runtime.

Our empirical evaluation is on benchmarks in C thus the choice of C++ to write mocks may seem un-intuitive.
However, C++ metaprogramming is the best possible solution in our opinion.
First, C++ enables a compile time DSL that is similar to GoogleMock but constructs a mock function at compile-time.
This cannot be accomplished using C macros alone to the best of our knowledge.
An alternative would be a custom processor for the DSL.
However, this would be complex since it would need to parse C code that may be contained in a DSL program.
Second, Using a DSL that allows dropping down to C++ enables full use of C/C++ expressiveness when needed, without the user learning new a language to be productive.
Third, though \seamock uses advanced metaprogramming, the user needs only a basic understanding of template metaprogramming.
The complexity is hidden behind an object oriented DSL.

\seamock, released as an open source library\footnote{\url{https://github.com/seahorn/seamock}}, is well suited for code-level verification tools that piggyback on a modern compilation strategy like \llvm~\cite{LLVM:CGO04}.
In our evaluation, we use \seahorn~\cite{DBLP:conf/fmcad/PriyaSBZVG22}, but expect \seamock to work with other verification tools such as \smack~\cite{DBLP:conf/cav/RakamaricE14}, \klee~\cite{DBLP:conf/osdi/CadarDE08}, or \crux~\cite{CruxUrl}.
\seamock is intended to be used in verification contexts but is not necessarily limited to it and we think it may be useful in unit testing as well.

To test the efficacy of \seamock, we use it to verify memory safety properties, which are important properties for security and reliability.
First, we evaluate how mocks differ from existing environment specification styles by writing different environments for the same unit proof that verifies a message handling function in the Android Trusty Trusted Execution Environment (TEE)~\cite{AndroidTrusty}.
 Second, we verify \verifiedfuncs public functions of the SSL messaging component of the \mbedtls~\cite{mbedTLS} library.
 This component has around $6\,000$ lines of code.
 We compare the unit proof development in \mbedtls with similar industrial verification projects~\cite{DBLP:journals/spe/ChongCEKKMSTTT21,KaniFirecracker} and find that, \emph{prima facie}, utilizing vMocks makes unit proof development at least three times faster.

In summary, the paper makes the following contributions:
\begin{inparaenum}[(1)]
  \item introduces the idea of using mocking (vMocks) in the context of code-level formal verification,
  \item a novel implementation of vMocks optimized for symbolic environments -- \seamock, and,
  \item illustration of efficacy of the vMock philosophy and the \seamock implementation through evaluation on an open source project.
\end{inparaenum}

The rest of the paper is organized as follows:
In \cref{sec:motivation}, we motivate vMocks by comparing its advantages to other environment styles in the Android Trusty project. In \cref{sec:seamock}, we present the design and implementation of \seamock library. In \cref{sec:evaluation}, we describe an extensive evaluation of vMocks on verification of memory safety of \mbedtls. We discuss usability issues in \cref{sec:use}, related work in \cref{sec:related}, and conclude in \cref{sec:conclusion}.
\ifarxivmode
\cref{app:background} provides background on mocking and how unit proofs are setup in verification.
\fi

\section{Motivation}
\label{sec:motivation}
To motivate the utility of mocks, we use our experience from verifying the Android Trusty Trusted Execution Environment (TEE)~\cite{AndroidTrusty} that provides a framework to build secure applications.
These applications must communicate with unsecure applications running outside the TEE.
The communication happens through an \code{ipc} layer.
\cref{fig:do_handle_msg} is a simplified representation of a core function \code{do_handle_msg} in the \code{ipc} layer. It reads an incoming message from the TEE kernel and calls an appropriate application function.
We want to verify that \code{do_handle_msg} has no out-of-bounds memory accesses and that it does not modify the message bytes before they are passed to the application.
These expectations are encoded in the unit proof in~\cref{fig:unit_proof_do_handle_msg}.
The environment consists of the following functions provided by the TEE.
Functions \code{create_channel} and \code{wait_for_msg} create a collection of channels and choose a channel non-deterministically, respectively.
The \code{get_msg} function returns the length of the incoming message.
The \code{read_msg} function copies the message contents into a provided buffer.
The \code{put_msg} function retires the message indicating that it is not accessed further by the application.
We use the \seahorn \bmc engine for verifying \code{do_handle_msg}.
Before looking at various styles for specifying the environment, we
provide some background on the verification tool itself.

\bparagraph{\seahorn.}
The \seahorn \bmc engine~\cite{DBLP:conf/fmcad/PriyaSBZVG22} is a bit precise bounded model checker for \llvm programs.
It uses \texttt{Clang} to compile C programs to \llvm bitcode.
It provides a number of builtins to enable writing specifications in the style of C functions.
\code{sassert} and \code{assume} codify verification assertions and assumptions respectively.
Undefined but declared functions starting with \code{nd_} return non deterministic values of the declared return type.
The \code{is_deref} builtin checks if a memory access is within allocated bounds.
The \seahorn pipeline automatically adds an \code{is_deref} check before every memory access.
The \code{is_modified} builtin checks if an allocation was modified since \code{reset_modified} was called on that allocation.
The \code{memhavoc} builtin havocs a newly created memory allocation.

\bparagraph{Unit proof.}
A verification task needs a harness to setup the environment and start the \sut with valid arguments passed to the chosen entry point.
A methodology for designing function specific harnesses is discussed in~\cite{DBLP:journals/spe/ChongCEKKMSTTT21}. In~\cite{DBLP:journals/isse/PriyaZSVBG22}, this harness specification is called a unit proof, after unit tests.
~\cref{fig:unit_proof_do_handle_msg} shows a typical unit proof.
The unit proof is packaged as a C \code{main} function.
Note that the specification language for pre-and-post conditions is C, the same language as the function under verification.
This is for ease of developers already familiar with the target language.
A detailed rationale is found in~\cite{DBLP:journals/isse/PriyaZSVBG22}, which calls it Code-As-Specification (CaS).
The setup involves creating two channels.
One of the channels is chosen using \code{wait_for_msg} at~\lref{line:unit_proof_wait_for_msg}.
Finally the \code{do_handle_msg} is called, passing in a message handler and the chosen channel.
In production, \code{do_handle_msg} passes the incoming message to the application using a passed-in message handler.
For the unit proof, the message handler is defined in \code{test_msg_handler}.
~\Lref{line:unit_proof_is_mod} checks that the given \code{msg} arg has not been modified since being reset.
The reset is part of the environment and hence not part of~\cref{fig:unit_proof_do_handle_msg}.
The \code{test_msg_handler} is free to return an \code{error} ($<0$) or \code{ok} ($>=0$) code.
Thus, we can return any non deterministic integer at~\lref{line:unit_proof_return_nd_int} using \code{nd_int}.

\bparagraph{Environment specification styles.}
A unit proof requires an environment to function correctly.
The environment, operationally, is a sequence of function calls that supply valid data to the \sut.
When it comes to modeling the \env, there are several design choices at hand. The \env can either be stateless, where it provides fresh, non--deterministic values with each function call, or stateful, in which case it holds a more comprehensive specification of the environment required for the SUT to function correctly. In the case of a stateless environment, each function call can be represented using a function summary, which is a declarative specification assuming a simple stateless environment.
However, if the environment requires state, it can be modeled using a construct referred to as a \emph{fake}.
We explore these ways of specifying the environment next.

\newsavebox{\figdohandlemsgbox}
\begin{lrbox}{\figdohandlemsgbox}%
	\begin{lstlisting}[style=SeaC, escapechar=@]
extern  int get_msg(int, size_t *);
extern  int read_msg(int, char *);
extern  int put_msg(int);

#define MAX_SIZE 4096
typedef int (*msg_handler_t)(char* msg,
                             size_t msg_size);
#define FREE_AND_RET(msg, rc) \
  do { free(msg); return rc; } while(0)
#define ERROR -1
int do_handle_msg(
    msg_handler_t msg_handler,
    int chan) {
  char *msg = (char *)malloc(MAX_SIZE);
  size_t msg_len;
  int rc = get_msg(chan, &msg_len);
  if (rc < 0) FREE_AND_RET(msg, ERROR);
  rc = read_msg(chan, msg);
  put_msg(chan);
  if (rc < 0) FREE_AND_RET(msg, ERROR);
  if (((size_t) rc) < msg_len)
    FREE_AND_RET(msg, ERROR); @\label{line:do_handle_msg_err3}@
  rc = msg_handler(msg, msg_len);
  FREE_AND_RET(msg, rc); 
}
\end{lstlisting}%
\end{lrbox}

\newsavebox{\figunitproofdohandlemsgbox}
\begin{lrbox}{\figunitproofdohandlemsgbox}%

	\begin{lstlisting}[style=SeaC, escapechar=@]
static int test_msg_handler(
    char *msg, size_t msg_size) {
  sassert(!sea_is_modified((char *)msg)); @\label{line:unit_proof_is_mod}@
  return nd_int();@\label{line:unit_proof_return_nd_int}@ }
int main(void) {
  // create 2 channels
  create_channel();
  create_channel(); 
  int chan = wait_for_msg(); @\label{line:unit_proof_wait_for_msg}@
  do_handle_msg(
      test_msg_handler, chan); @\label{line:unit_proof_do_handle_msg_call}@
  return 0; 
}
\end{lstlisting}%
\end{lrbox}


\begin{figure}[t]
	\subcaptionbox{C function under verification.\label{fig:do_handle_msg}}[0.40\linewidth]{%
		{\scalebox{1.0}{\usebox{\figdohandlemsgbox}}}}
	\hspace{1cm}
	\subcaptionbox{Unit proof for \code{do_handle_msg}.\label{fig:unit_proof_do_handle_msg}}[0.5\linewidth]{%
		{\scalebox{1.0}{\usebox{\figunitproofdohandlemsgbox}}}}
	\caption{An \sut and its unit proof from Android Trusty TEE.}
	\label{fig:sut_and_unit_proof}
	\vspace{-0.5cm}
\end{figure}

\newsavebox{\figfakeenvbox}
\begin{lrbox}{\figfakeenvbox}%

	\begin{lstlisting}[style=SeaC, escapechar=@, numbers=left]
#define MAX_CHANNELS 10

// The next channel number to initialize
// Incremented on every create_channel call
static int g_next_available_channel;
// The first yet unprocessed channel
// Incremented on every put_msg call
static int g_already_processed_channel;

typedef struct msg {
  char* buf;
  size_t len;
} MSG;

static MSG msgs[MAX_CHANNELS];

int create_channel() {@\label{line:fake_example_create_channel}@
  sassert(g_next_available_channel <
    MAX_CHANNELS);
  size_t len = nd_size_t();@\label{line:fake_example_create_channel_msg_len}@
  char *msg = (char *)malloc(len);
  memhavoc(msg, len);@\label{fake_example_memhavoc}@
  int chan = g_next_available_channel++;
  msgs[chan].buf = msg;
  msgs[chan].len = len;
  return chan;}

int wait_for_msg() {@\label{line:fake_example_wait_for_msg}@
  int channel = nd_int();
  assume(
    channel > g_already_processed_channel &&
    channel < g_next_available_channel);
  return channel;}

int get_msg(int chan, size_t *len) {@\label{line:fake_example_get_msg}@
  int err_code = nd_int();
  if (err_code < 0) return err_code; @\label{line:fake_example_get_msg_ret_err}@
  sassert(chan > 0 &&
      chan < MAX_CHANNELS &&
      len != NULL);
 *len =  msgs[chan].len;
  return 0; }

int read_msg(int chan, char *msg) {@\label{line:fake_example_read_msg}@
  sassert(chan > 0 &&
      chan < g_next_available_channel && 
      msg != NULL);
  memcpy(msg,  msgs[chan].buf,  msgs[chan].len);
  sea_reset_modified(msg);
  return 0;}

int put_msg(int chan) {@\label{line:fake_example_put_msg}@
  sassert(chan > 0 && chan <
      g_next_available_channel);
  int err_code = nd_int();
  if (err_code < 0) return err_code;
  free(msgs[chan].buf);
  g_already_processed_channel++;
  return 0;}
\end{lstlisting}%
\end{lrbox}%

\newsavebox{\figsummenvbox}
\begin{lrbox}{\figsummenvbox}%
	\begin{lstlisting}[style=SeaC, escapechar=@, numbers=left]
#define MAX_SIZE 4096

int create_channel() {/* NOP */ return nd_int(); }
int wait_for_msg() {/* NOP */ return nd_int(); }

int get_msg(int chan, size_t *len) {
  sassert(chan > 0 && len != NULL);
  *len = nd_size_t();
  return nd_int(); // error_code  
}

int read_msg(int chan, char *msg) {
  sassert(chan > 0 && msg != NULL);
  char buf[MAX_SIZE];
  memhavoc(&buf);
  size_t len = nd_size_t();  @\label{line:read_msg_summary_err1}@
  assume(len < MAX_SIZE); 
  memcpy(msg, buf, len);
  sea_reset_modified(msg);
  return nd_bool() ? -1 : len; }

int put_msg(int chan) {
  sassert(chan > 0);
  return nd_int(); }
\end{lstlisting}%
\end{lrbox}%

\newsavebox{\figmockenvbox}
\begin{lrbox}{\figmockenvbox}%

	\begin{lstlisting}[style=SeaC, escapechar=@, numbers=left]
#define MAX_SIZE 4096

int create_channel() {/* NOP */ return nd_int(); }
int wait_for_msg() {/* NOP */ return nd_int(); }

static size_t g_msg_size;

int get_msg(int chan, size_t *len) {
  sassert(chan > 0 && len != NULL);
  g_msg_size = nd_size_t();
  return nd_int(); }

int read_msg(int chan, char *msg) {
  sassert(chan > 0 && msg != NULL);
  char buf[MAX_SIZE];
  memhavoc(&buf);
  size_t len = nd_size_t();
  assume(len <= g_msg_size);  @\label{line:read_msg_mock_assume}@
  memcpy(msg, buf, len);
  sea_reset_modified(msg);
  return len; }

int put_msg(int chan) {
  sassert(chan > 0);
  return nd_int(); }
\end{lstlisting}%
\end{lrbox}%

\begin{figure}[t]
	\centering
	\begin{minipage}[t]{0.45\linewidth}
		\subcaptionbox{A fake environment.\label{fig:fake_environment}}[0.9\linewidth]{%
			\scalebox{1.0}{\usebox{\figfakeenvbox}}}
	\end{minipage}%
	\hfill
	\begin{minipage}[t]{0.45\linewidth}
		\raisebox{10cm}{\subcaptionbox{Function summaries for environment.\label{fig:function_summaries}}[0.9\linewidth]{%
				{\scalebox{1.0}{\usebox{\figsummenvbox}}}}} \\[-9cm]
		\subcaptionbox{A mock environment.\label{fig:mock_environment}}[0.9\linewidth]{%
			\scalebox{1.0}{\usebox{\figmockenvbox}}}
	\end{minipage}
	\caption{Different environment specification styles for \code{do_handle_msg}.}
\end{figure}%

\bparagraph{Option1: Faking the environment.}
The most general approach to modelling the environment is by creating a fake -- a replica of the original environment with simpler internals.
A fake environment for \code{do_handle_msg} is defined in~\cref{fig:fake_environment}.
This environment fakes incoming messages by providing an array of messages indexed by a channel.

To simulate multiple clients connecting to the TEE, the function \code{create_channel} creates a channel with a pending message and adds it to a global queue of messages \code{msgs}.
~\Lref{line:fake_example_create_channel_msg_len} chooses a non--deterministic length for the message.
The \code{memhavoc} builtin at~\lref{fake_example_memhavoc} marks the message buffer as containing non--deterministic sequence of bytes.
\code{memhavoc} ensures that \code{do_handle_msg} does not depend on a specific byte sequence, especially, if we want to check that a message is never altered by the function.
The function \code{wait_for_msg} chooses from channels available currently.
Thus, when \code{create_channel} function is called from the unit proof in~\cref{fig:unit_proof_do_handle_msg}, it choose from one of two channels.

The \code{get_msg} function may return an error so the return value is non--deterministic at~\lref{line:fake_example_get_msg_ret_err}.
The subsequent \code{sassert} statement checks that the function pre-conditions are met.

The \code{read_msg} function checks pre-conditions and copies the (non--deterministic) message from the internal message buffer to the one passed in, i.e. \code{msg}.
It also invokes a builtin \code{sea_reset_modified} to mark \code{msg} as being sensitive to modification.
Henceforth, any store to \code{msg} will taint the memory and cause the corresponding \code{sea_is_modified} builtin to return true.

Finally, the function \code{put_msg} retires the message.
In our simple fake, we only want a channel to provide a single message so we mark the channel as processed using a counter \code{g_already_processed_channel}.

Though a fake faithfully represents the actual environment, there are two problems that arise in using them in verification.
First, the fake may require complicated internal data structures that are expensive to symbolically execute (in a \bmc tool).
In this example, the choice of the array data structure is important since it is efficient.
A simple associative array data structure suffices because a channel is really an opaque type \code{handle_t}.
Since it is opaque, we are free to map it to \code{int} and use an array instead of a more expensive data structure.
Second, fakes themselves may become complex enough to require their own verification.
This adds to the verification burden of the project.
To circumvent the problems posed by fakes, we next look at how a minimum environment can be modelled using function summaries.

\bparagraph{Option 2: Function summaries.}
A stateful environment is complex.
A stateless environment is easier to model.
\Cref{fig:function_summaries} defines the summaries for our running example.
We use function summaries to (1) declare the preconditions for each function in the environment and, (2) to declare valid values that the function can output through output arguments and return statements.
Notice that we don't need to provide any summary for \code{create_channel} and \code{wait_for_msg} since these functions are never called from \code{do_handle_msg}.
Unfortunately, this summary specification is not strong enough to guarantee correctness of \code{do_handle_msg}.
The summary of \code{read_msg} chooses a non--deterministic integer for \code{len} in~\lref{line:read_msg_summary_err1}.
The length maybe greater than the size of \code{msg} buffer.
This condition is not handled in \code{do_handle_msg}.
It also leads to undefined behaviour in \code{memcpy}.
Thus, we discover that a correctly functioning environment has a stronger specification -- that the length returned by \code{get_msg} is less--than--or--equal to that used in \code{read_msg}.

\bparagraph{Towards mock environments.}
We have learned a valuable lesson from the exercise so far.
On the one hand, a fake environment works in practice but is expensive to construct.
\otoh, using function summaries enables us to create an inexpensive partial environment but it may need minimum state to satisfy client requirements.
Thus, we want a partial environment that is just right -- with function summaries that encourage non--determinism and minimum state for the correct functioning of the program function.

For our example,~\cref{fig:mock_environment} shows a partial environment that records the length when \code{get_msg} is called and recalls it for \code{read_msg} at~\lref{line:read_msg_mock_assume}.
It is similar to~\cref{fig:function_summaries} in other respects.
This satisfies our immediate goals.
However, on closer examination, we see that we have really designed the new environment in response to how it is to \emph{behave} with \code{do_handle_msg}.
In particular, it implicitly assumes that \code{get_msg} is called before \code{read_msg} so that message length (\code{g_msg_size}) is set before it is read.
In concrete testing, such behavorial specifications are codified using tMocks.
We thus want their counterpart in verification i.e., vMocks.
We also note a key difference between a \fak and a mock.


\bparagraph{Discussion.}
We summarise our experience in going from a fake to a mock.
Developing a fake required the following steps:
\begin{inparaenum}[(1)]
	\item understanding the \code{ipc} layer, corresponding environment interface and implicit contracts between the \sut and environment,
	\item implementing a fake that is faithful to external contracts but internally simple for verification, and,
	\item developing unit proofs for the fake, finding and fixing bugs.
\end{inparaenum}
Thus, developing a fake became its own project with the usual develop-debug-fix cycle.
In contrast, we developed a vMock environment just for \code{do_handle_msg}.
Thus, it was simple enough to not require verification.
Additionally, with \seamock, the implementation was auto-generated, removing the opportunity for trivial errors.
To make vMocks practical, we require a DSL that satisfies two criteria.
First, it must be similar to how mocks are specified in concrete tests.
This would make it attractive to developers who already have experience with mocks.
Second, it must be inexpensive to execute symbolically (e.g., in a \bmc tool).
Thus, whatever language constructs it provides, must produce minimum overhead during verification.
To satisfy these requirements, we describe the \seamock framework for creating user friendly and efficient vMocks in the next section.

\afterpage{\clearpage} 
\section{The \seamock framework}
\label{sec:seamock}

\begin{table}[t]
	\scriptsize
	\centering
	\rowcolors{2}{gray!30}{white} 
	\begin{tabular}{ l|l }
		Expectation (method)             & Meaning                                                                                 \\
		\hline
		\code{times(Predicate<N>())}     & \text{Predicate ($<,>,=$) applied to number of mock calls and \textit{N} is satisfied.} \\
		\code{returnFn(f)}               & \text{Mock function returns output of function \textit{f}.}                             \\
		\code{captureArgAndInvoke<N>(f)} & \text{Mock function captures $N$th positional argument, applies function $f$ on it.}    \\
		\code{invokeFn(f)}               & \text{Mock function invokes function \textit{f} with all arguments.}
	\end{tabular}
	\caption{vMock expectation DSL.}
	\label{tab:vmock_lang}
	\vspace{-0.5cm}
\end{table}%
\bparagraph{Requirements analysis.}
tMock frameworks are designed to be executed in concrete environments.
The same design choices may not be ideal for symbolic environments.
For example, in gMock, calls to mock functions are resolved using dynamic dispatch.
Resolving such calls dynamically in a symbolic environment is, in general, hard.
Another framework, cMocka~\cite{CMocka} for C programs aims to have minimal dependencies on libraries or latest compiler features for wide applicability.
It uses a runtime map keyed by function names to store expected behaviours.
This runtime map is similarly expensive to use in a symbolic environment.

In some cases, it may be possible to adapt symbolic environments to work well with existing mock frameworks.
However, we take a novel approach of creating a gMock inspired framework and DSL where function calls are resolved at compile time while retaining the familiar call-chain syntax.
This is done using modern C++ template meta-programming\footnote{\seamock is built against C++17.}. The framework is open sourced at~\url{https://github.com/seahorn/seamock}.
We use C++ meta-programming instead of a custom preprocessor because the template mechanism is supported by industrial C++ compilers is widely available and avoids dependencies on tools that may not be available on a particular platform.

It is important to note that \seamock does not aim to be a general purpose mocking framework.
It may be useful to adapt it to unit testing, however the current implementation depends on a compiler with modern C++ features and has been tested on the \texttt{Clang} compiler toolchain only. We do expect \seamock to work with other verification tools such as \smack~\cite{DBLP:conf/cav/RakamaricE14}, \klee~\cite{DBLP:conf/osdi/CadarDE08}, or \crux~\cite{CruxUrl} that use \llvm.

\bparagraph{Usage.}
The \seamock DSL is embedded in C++.
It is in the spirit of gMock which in-turn is inspired by jMock~\cite{jMockUrl} and EasyMock~\cite{EasyMockUrl}.
jMock and gMock use a declarative scheme to build expectations on a mock object using a call chain syntax (builder pattern) ~\cite{DBLP:conf/oopsla/FreemanP06}.
The declarative scheme to specify tMock behavior is well established in industry and thus familiar to developers.
Therefore \seamock DSL also uses a declarative builder scheme as shown in~\cref{tab:vmock_lang}.
An \code{ExpectationBuilder} constructs an expectation object using methods to setup expectations before calling \code{build()}.
For example, the method call \code{times(Eq<3>())} creates an expectation that a mock function must be called \code{3} times in all execution paths.
A built expectation can be attached to a mock function using
\code{MOCK_FUNCTION(name, expectation, return_type, (argument_type, ...))} where \code{name} is the function name, \code{expectation} is the user expectation, \code{return_type} is the return type of the function, and (\code{argument_type,...}) a  tuple of arguments for the function.
To constraint order of execution, the macro \code{MOCK_FUNCTION_W_ORDER(..., (MAKE_PRED_FN SET(predecessor_fn)), ...)} is used.

\newsavebox{\figvmockbox}
\begin{lrbox}{\figvmockbox}%
	\begin{lstlisting}[style=SeaC++, escapechar=@, numbers=left]
static size_t g_msg_size;
// *** Begin: Mock arg capture ***@\label{line:vmock-begin-arg}@
static constexpr auto set_ptr_fn_get_msg = [](size_t *len) {@\label{line:vmock-get-msg-capture-fn}@
  *len = nd_size_t();
  g_msg_size = *len;
};
static constexpr auto set_ptr_fn_read_msg = [](char *msg) {
  char *blob = (char *)malloc(g_msg_size);
  memhavoc(blob, g_msg_size);
  sassert(msg);
  memcpy((size_t *)msg, (size_t *)blob, g_msg_size);
  sea_reset_modified(msg);
};
// *** End: Mock arg capture ***
// *** Begin: mock expect definition ***
constexpr auto get_msg_expectations = ExpectationBuilder() @\label{line:vmock-get-msg-expect-start}@
                                              .times(Eq<1>())
                                              .returnFn(nd_int)@\label{line:vmock-get-msg-return-fn}@
                                              .captureArgAndInvoke<1>(set_ptr_fn_get_msg)@\label{line:vmock-get-msg-capture-set}@
                                              .build();@\label{line:vmock-get-msg-expect-end}@

constexpr auto read_msg_expectations = ExpectationBuilder()
                                              .times(Lt<2>())
                                              .returnFn(MOCK_UTIL_WRAP_VAL(g_msg_size))
                                              .captureArgAndInvoke<1>(set_ptr_fn_read_msg)
                                              .build();
// *** End: mock expect definition ***
extern "C" {
MOCK_FUNCTION(get_msg, get_msg_expectations, int, (int, size_t *))@\label{line:vmock-get-msg-mock-def}@
MOCK_FUNCTION_W_ORDER(read_msg, read_msg_expectations, MAKE_PRED_FN_SET(get_msg), int, (int, char *))@\label{line:vmock-read-msg-expect-after}@
LAZY_MOCK_FUNCTION(put_msg, int, (int))@\label{line:vmock-put-msg-mock-def}@
SETUP_POST_CHECKS((get_msg, read_msg, put_msg))}
\end{lstlisting}%
\end{lrbox}%
\begin{figure}[t]
	\centering
	\scalebox{1.0}{\usebox{\figvmockbox}}
	\caption{vMock example using \seamock.}
	\label{fig:vmock_example}
	\vspace{-0.5cm}
\end{figure}%
\newsavebox{\figsemanticsone}
\begin{lrbox}{\figsemanticsone}%
	\begin{minipage}{\textwidth}
		{\scriptsize
			\begin{align*}
				insert(M, k ,v)       & \sdef M \cup {(k,v)} \textit{, where } M : \cvtype \rightarrow \cvtype, k:\cvtype,v:\cvtype                        \\
				atkey(M, k)           & \sdef v \text{ if } (k,v) \in M \textit{, where } M : \cvtype \rightarrow \cvtype,k:\cvtype,v:\cvtype              \\
				putReturnFn(E, retfn) & \sdef insert(M, ReturnFn, retfn) \textit{, where } E : \cstype \rightarrow \cvtype,ReturnFn:\cstype, retfn:\cftype \\
				getReturnFn(E)        & \sdef atkey(E, ReturnFn) \textit{, where } E : \cstype \rightarrow \cvtype, ReturnFn:\cstype                       \\
				skeletal(E)           & \sdef  getReturnFn(E)()
			\end{align*}
		}
	\end{minipage}
\end{lrbox}%

\newsavebox{\figsemanticstwo}
\begin{lrbox}{\figsemanticstwo}%
	\begin{minipage}{\textwidth}
		{\scriptsize
			\begin{align*}
				putCaptureMap(E, capmap)       & \sdef insert(E, CaptureMap, capmap) \textit{, where } E : \cstype \rightarrow \cvtype,          & \\& \qquad CaptureMap:\cstype, capmap:\cntype \rightarrow \cftype &\\
				getCaptureMap(E)               & \sdef atkey(E, CaptureMap) \textit{, where } E : \cstype \rightarrow \cvtype,CaptureMap:\cstype & \\
				args                           & \sdef \langle arg_{1}, arg_{2}, ...,\arg_{n} \rangle                                            & \\
				invokeWithCapturedArg(args, E) & \sdef \{ f_j(args[i_j]) \mid (i_j, f_j) \in getCaptureMap(E) \}                                 &
			\end{align*}
		}
	\end{minipage}
\end{lrbox}%

\begin{figure}[t]
	\centering
	\scalebox{1.0}{\usebox{\figsemanticsone}}
	\caption{Expectation map (M) definition and mechanism for wiring a return function.}
	\label{fig:expect_map_defs}
	\vspace{-0.8cm}
\end{figure}
\begin{figure}[t]
	\centering
	\scalebox{1.0}{\usebox{\figsemanticstwo}}
	\caption{High-level mechanism for \texttt{captureArgAndInvoke}.}
	\label{fig:capture_arg_defs}
	\vspace{-0.8cm}
\end{figure}
In~\cref{fig:vmock_example},~\lref{line:vmock-get-msg-mock-def} shows how the mock \code{get_msg} function is specified.
The expectation is built above in~\lref{line:vmock-get-msg-expect-start} to \lref{line:vmock-get-msg-expect-end}.
The \code{ReturnFn} for \code{get_msg} is set in~\lref{line:vmock-get-msg-return-fn}.
It returns a non--deterministic \code{size_t} value.
We would like to capture the pointer \code{len} and set it to a non--deterministic value. This is the first (starting from zero) argument for \code{get_msg}.
This is setup in two steps.
First, the effect of the capture is defined using a lambda at~\lref{line:vmock-get-msg-capture-fn}.
Second, the lambda is wired to the correct positional argument using a setter at~\lref{line:vmock-get-msg-capture-set}.
\seamock provides built--in logic like call cardinality and sequencing that have to be hand--coded otherwise.

In the spirit of need driven development~\cite{DBLP:conf/oopsla/FreemanMPW04a}, the user only has to specify the absolute minimum required. If a particular expectation is unspecified then a mock function is assembled with defaults.
For example, we do not specify the order constraint in~\cref{fig:vmock_example},~\lref{line:vmock-get-msg-mock-def} since we don't expect any other mock function to be called before \code{get_msg}.  The default (no expectation) is assumed when a particular constraint is unspecified.
We do specify an order for the \code{read_msg} mock in~\lref{line:vmock-read-msg-expect-after} because we expect that if it is called in an execution, then it is always after \code{get_msg}.
Last, if we just want to specify a function summary -- that just returns non--deterministic values then a mock can be declared lazy using \code{LAZY_MOCK_FUNCTION} with no expectations specified as in~\lref{line:vmock-put-msg-mock-def}.


\bparagraph{Implementation.}
\seamock is implemented using C++ metaprogramming.
The metaprogram constructs mock functions from the expectations setup through the DSL.
It is developed using the Boost Hana library~\cite{BoostHana}, which provides a functional programming layer over the base metaprogramming environment.
Before explaining the details particular to C++, we discuss the high level design of the library.
Each mock function requires a mapping of an expectation to an action. For example, a return function expectation results in returning the value of a particular function application on mock function exit.
This mapping, called the expectation map, is setup by the user using the \seamock DSL.
The map is a compile-time entity and thus after the C++ template pre-processor executes,
a mock function is wired with values gotten from its expectation map and the map ceases to exist.
Specifically, there is no lookup at run-time.

We present an example of how a return function expectation is setup internally in the framework in~\cref{fig:expect_map_defs} using a compile-time expectation map (E).
Basic types available to the compile time metaprogram are constant naturals ($\cntype$), constant string ($\cstype$) and a function object ($\cftype$). These collectively represent constant value types ($\cvtype$).
The expectation map stores an action (data) keyed by an expectation, which is of type $\cstype$.
The $putReturnFn$ function inserts a value (a 0--arity function object) for the $ReturnFn$ key.
The $getReturnFn$ function  extracts an inserted function object for the $ReturnFn$ key.
The $getReturnFn$ function is applied in a $skeletal$ function, that is the core of a mock function.

In~\cref{fig:capture_arg_defs}, the mechanism of \code{captureArgAndInvoke<N>} is defined.
Through the DSL, the user provides a mapping ($CaptureMap$) from function argument positions to invoked functions.
The framework constructs a tuple ($args$) from the given function arguments.
The $skeletal$ uses $CaptureMap$ and $args$ to match functions to arguments and apply the functions left-to-right.

The Boost Hana functional C++ metaprogramming library is used instead of the bare-bones template metaprogramming environment to improve programmer efficiency and correctness since C++ metaprogramming can get tedious and error-prone to work with in advanced use cases.
Note that all Hana abstractions are compiled away and thus incur no runtime cost.
Hana requires C++14 or beyond.
\seamock uses C++17 to allow the use of \code{constexpr} lambdas, which are not necessarily used currently but maybe useful in future.
Only the mock environment needs a C++ compiler.
Once this unit is compiled to \llvm IR, it can be linked to the \sut written in C/C++ that has also been compiled to \llvm IR.
Thus, the choice of implementing \seamock in C++ generally does not limit the \sut source language and C may be used.
An exception to this rule is when a C header file contains a function definition that may be accepted by the C compiler but not the C++ compiler.
We found this case during evaluation and discuss it in~\cref{sec:evaluation}.

A simplified low-level implementation of the \seamock internals is shown in~\cref{fig:seamock_hana_impl}.
The core of the \seamock implementation is the \code{skeletal} lambda
function at~\lref{line:seamock-skeletal}.
The \code{skeletal} function creates a mock function with desired actions at compile time based on the expectations setup by the user.
The user sets up expectations using the \code{MOCK_FUNCTION} macro.
For simplicity, only a return function expectation is setup.
A mock function is constructed in the following way at compile time.
\begin{inparaenum}
	\item[(1)] A return function is added to the expectation map at~\cref{line:seamock-return-fn-wire} through the user DSL.
	\item[(2)] A mock function definition is created at~\lref{line:seamock-mock-fn-def}.
	\item[(3)] The definition applies the \code{skeletal} function to the
	expectation map and the arguments passed to the mock function at~\lref{line:seamock-apply-skeletal}.
	This application creates a specialized \code{skeletal} lambda for the given mock function.
\end{inparaenum}

Developers use tMocks through a fluent interface (builder like pattern).
To enable easy adoption of vMocks, we encapsulate the metaprogramming in the same pattern.
Specifically \seamock uses a compile-time builder pattern (available in C++11 and beyond).
A simplified version is shown in~\cref{fig:seamock_oop_impl}.
Each call to a setter method adds a key,value pair to the expectations map.
The fully constructed expectation maps is returned using the \code{build} method.
In summary, \seamock provides a familiar interface to developers and they need not know advanced C++ metaprogramming or the Boost Hana framework to be productive. In the next section, we describe our experience in using \seamock to verify an open source industrial project, \mbedtls.
\newsavebox{\figseamockbox}
\begin{lrbox}{\figseamockbox}%
	\begin{lstlisting}[style=SeaC++, escapechar=@, numbers=left,xleftmargin=.2\textwidth]
constexpr auto ReturnFn = [](auto ret_fn_val, auto expectations_map) {@\label{line:seamock-return-fn-wire}@
  auto tmp = hana::erase_key(expectations_map, RETURN_FN);
  return hana::insert(tmp, hana::make_pair(RETURN_FN, ret_fn_val));
};
// Use: idx=i, type=T1 -> T1 argi
#define CREATE_PARAM(r, data, idx, type) (type BOOST_PP_CAT(arg, idx))
// Use: (int, float, char) ->
//     int arg0, float arg1, char arg2
#define UNPACK_TRANSFORM_TUPLE(func, tuple)  \
  BOOST_PP_TUPLE_ENUM(BOOST_PP_SEQ_TO_TUPLE( \
  BOOST_PP_SEQ_FOR_EACH_I(func, DONT_CARE,   \
      BOOST_PP_TUPLE_TO_SEQ(tuple))))
// Expectation map key
#define RETURN_FN BOOST_HANA_STRING("return_fn")
static constexpr auto Defaults = hana::make_map();
#define CREATE_ND_FUNC_NAME(name, type)  \
    BOOST_PP_CAT(nd_, \
    BOOST_PP_CAT(name, BOOST_PP_CAT(type, _fn)))

#define MOCK_FUNCTION(name,                  \
    expect_map, ret_type, args_tuple)        \
    ret_type name(                           \
    UNPACK_TRANSFORM_TUPLE(CREATE_PARAM,     \
        args_tuple)) {@\label{line:seamock-mock-fn-def}@\
    return hana::apply(skeletal, expect_map,@\label{line:seamock-apply-skeletal}@\
        hana::make_tuple(                    \
        UNPACK_TRANSFORM_TUPLE(CREATE_ARG, args_tuple))); \
  }
static auto skeletal = [](auto &&expectations_map,@\label{line:seamock-skeletal}@
    auto &&args_tuple) {
  auto ret_fn = hana::at_key(expectations_map,
      RETURN_FN);
  return ret_fn();
};
\end{lstlisting}%
\end{lrbox}

\newsavebox{\figseamockoopbox}
\begin{lrbox}{\figseamockoopbox}%
	\begin{lstlisting}[style=SeaC++, escapechar=@, numbers=left,xleftmargin=.2\textwidth]
    template
    <typename MapType=decltype(DefaultExpectationsMap)>
class ExpectationBuilder {
private:
    MapType expectationsMap;
public:
    // Constructor to initialize with an existing map
    constexpr ExpectationBuilder(
        const MapType& map) : expectationsMap(map) {}

    // Default constructor
    constexpr ExpectationBuilder() :
        expectationsMap(DefaultExpectationsMap) {}

    template<typename ReturnFnType>
    constexpr auto returnFn(ReturnFnType i) const {
      auto updatedMap = ReturnFn(i, expectationsMap);
      return ExpectationBuilder<
          decltype(updatedMap)>(updatedMap);
    }

    // Finalize build
    constexpr auto build() const {
        return expectationsMap;
    }
};

  \end{lstlisting}%
\end{lrbox}

\begin{figure}[t]
	\begin{subfigure}[b]{0.57\linewidth}
		\scalebox{1.0}{\usebox{\figseamockbox}}
		\caption{Wiring of Simplified \code{MOCK_FUNCTION} implementation.}
		\label{fig:seamock_hana_impl}
	\end{subfigure}
	\begin{subfigure}[b]{0.4\linewidth}
		\scalebox{1.0}{\usebox{\figseamockoopbox}}
		\caption{Expectation Builder.}
		\label{fig:seamock_oop_impl}
	\end{subfigure}
	\caption{Seamock implementation. User interacts using the Builder interface (~\cref{fig:seamock_oop_impl}).}
	\label{fig:seamock_impl}
	\vspace{-0.8cm}
\end{figure}%

\section{Evaluation}
\label{sec:evaluation}
\begin{figure}[t]
  \centering
  \begin{adjustbox}{margin=-1cm 0 0 0}
    \includegraphics[trim={0cm 0.5cm 0cm 0cm},clip,scale=0.22]{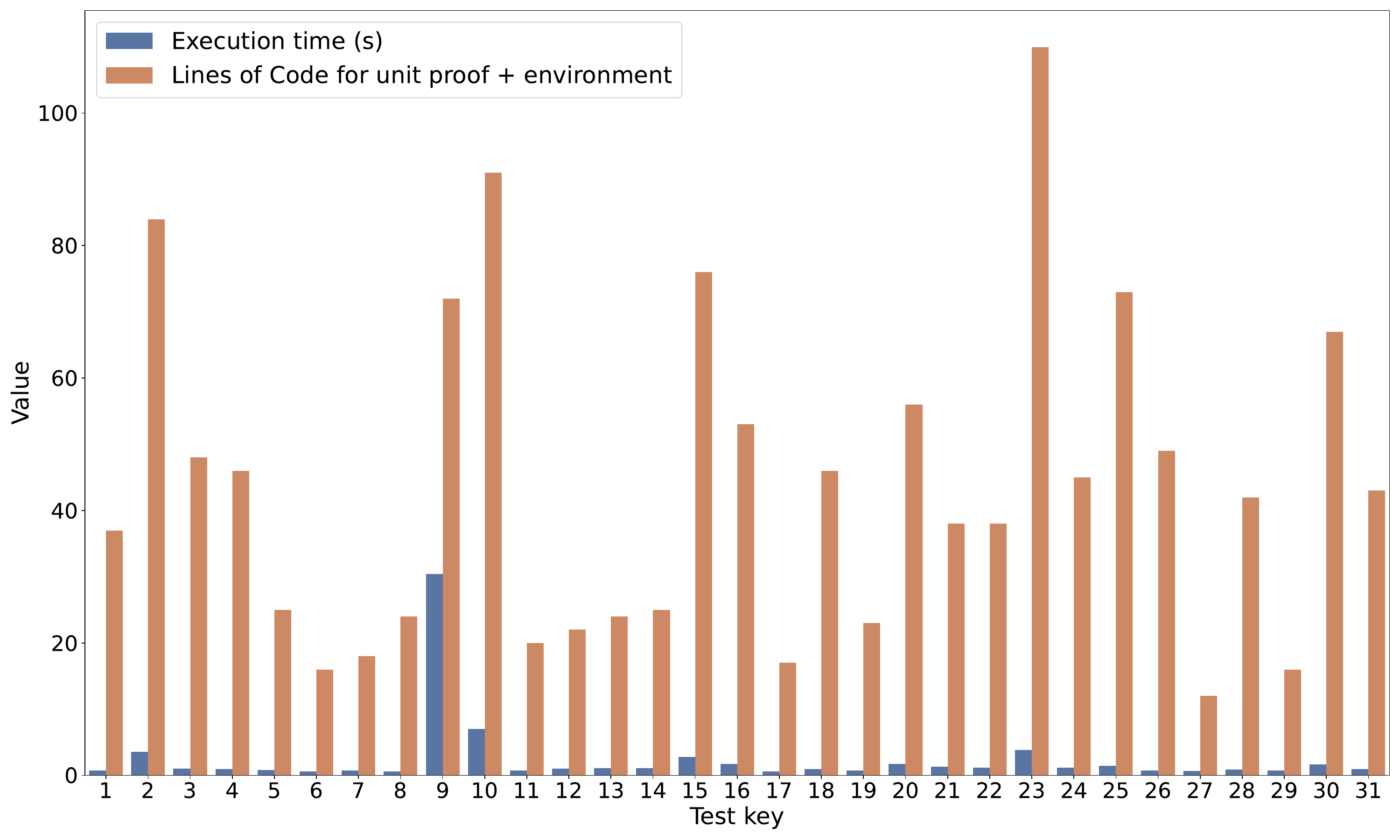}
  \end{adjustbox}
\ifarxivmode
  \caption{Performance and size of unit proofs. Key to test name map in~\cref{app:vmock-appendix},~\cref{tab:key_to_testname}.}
\else
  \caption{Performance and size of unit proofs. Key to test name map in the extended version~\cite{Priya2024vMocksArxiv}.}
\fi
\label{fig:mbedtls_perf}
\vspace{-0.9cm}
\end{figure}

To evaluate vMocks, we verified a core component of the \mbedtls open source project using \seamock and \seahorn.
The \mbedtls project is a C library that implements cryptographic primitives, SSL/TLS and DTLS protocols.
Its small code footprint and multiple configuration options makes it suitable for embedded systems.
The choice of \mbedtls was guided by our industrial partners, who use the library extensively in their codebases.

In \mbedtls, we undertook the exercise of verifying functions dealing with SSL messages.
These functions are present in the \code{ssl_msg.c} source file which is $6\,000$ lines of code including comments.
The goal of this exercise was to understand the tradeoffs provided by vMocks in formal verification.
Thus, our aim was \emph{not} to find bugs in the implementation or prove their absence.
Our goals were to study the following:
\begin{inparaenum}[(1)]
\item The practicality of developing unit proofs for a  \emph{legacy} codebase (without unit tests),
\item the flexibility of using \seamock to develop environments, and,
\item a qualitative understanding of how code structure affects development of unit proofs.
\end{inparaenum}

The \code{ssl_msg.c} compilation unit contains \totalfuncs public (non-static) functions.
For our evaluation, we decided to verify low level properties, specially memory safety for these public functions.
We developed unit proofs for \verifiedfuncs functions.
Of the remaining \notverifiedfuncs public functions, \trivialfuncs functions are trivial -- getters, setters and simple boolean checks.
The remaining \notinterestedfuncs either were specifications themselves or did not have memory accesses. See%
\ifarxivmode
  ~\cref{app:vmock-appendix}
\else
  the extended version~\cite{Priya2024vMocksArxiv}
\fi
for details.
We were able to develop unit proofs utilizing vMocks for all functions of interest.
Almost all unit proofs used the \seamock DSL, the remaining few had to be hand coded in C because their environments needed a header file (\code{ssl_misc.h}) that compiled under a C compiler but not under a C++ compiler.
\seamock can be used for all functions once this problem is corrected at source in the \mbedtls project.
The experiments were conducted on an Intel(R) Xeon(R) CPU E5-2680 running at 2.70GHz with 64 GiB of main memory. Total build time for all unit proofs was 5 seconds.

The results are given in~\cref{fig:mbedtls_perf}.
All execution times are single seconds except \code{mbedtls_ssl_read} (key=9) where the unit proof checks that all bytes are copied from one buffer to another. This is a more involved proof and takes closer to a half a minute to run.
Of particular interest is the source code complexity for the unit proof and the environment.
The average unit proof (harness) size is 24 lines of code and the average environment size is 47 lines of code.
Both excluding comments.
This is an indication that developing a single unit proof does not involve a lot of coding.
The artifacts from the study are open sourced~\cite{priya_gurfinkel_kahsai_2024}. 
We discuss our findings next.

\bparagraph{Process.}
Here we describe our process of verifying the SSL message component of the \mbedtls library.
First, useful pre-and-post conditions were intuited by breaking the overall verification goal into unit proofs aimed at verifying individual functions.
Next, using a list of callees of the function, we decided on functions to mock --
we sometimes mocked only a subset of callees to verify more of the production code.
Lastly, a decision was made on how \emph{lazy} a mock function should be.
Implementations written using \code{LAZY_MOCK_FUNCTION} did not add any expectations on the mock function.
For verifying memory safety, if the production function accessed passed-in buffers, then the mock function checked that such buffers were large enough to accommodate all accesses.

\bparagraph{The vMock language.}
We found that in verifying a legacy codebase (without unit tests) written by a third party was challenging because the pre-and-post conditions were not made explicit and the verification engineer could not explicate what these invariants could be.
Thus we found that coming up with the declarative behaviour of an environment was complex.
Instead, it was practical to use the \code{InvokeFn} action to invoke a simple function stub.
Thus, most environments were written as such rather than using more behaviorial actions like \code{After} and \code{ReturnFn}.
We posit that in a new software project, a rich set of actions in the mocking DSL is more useful since the environment is yet unspecified and behaviors are easier to model.

A typical example of a mock function is shown in \cref{fig:fn_ssl_recv}.
Here we see the use of \code{InvokeFn} to invoke a function at line \ref{line:invoke_fn_recv}.
The function checks that the passed-in \code{buf} can be dereferenced upto \code{len} bytes.
We also want to provide at least as many bytes as requested by the unit proof.
To wire this logic in, the unit proof calls the \code{set_min_recv_bytes} function before calling the \sut.
Now when \code{ssl_recv_fn} is called, we make sure that atleast that many bytes are returned.

\newsavebox{\figmbedtlsbox}
\begin{lrbox}{\figmbedtlsbox}%
\begin{lstlisting}[style=SeaC++, escapechar=@]
static size_t nb_bytes;

constexpr auto invoke_fn_mbedtls_ssl_recv_t = []@\label{line:invoke_fn_recv}@
(void *ctx, unsigned char *buf, size_t len) {
  if (buf != NULL) {
    sassert(sea_is_deref(buf, len));
  }
  int ret = nd_int();
  assume(ret <= 0 || ret >= (int)nb_bytes);
  return ret;
};
extern "C" {
void set_min_recv_bytes(size_t num_bytes) { nb_bytes = num_bytes; }
constexpr auto expectations_mbedtls_ssl_recv_t =
    seamock::ExpectationBuilder()
        .times(seamock::Lt<2>())
        .invokeFn(invoke_fn_mbedtls_ssl_recv_t)
        .build();
MOCK_FUNCTION(ssl_recv_fn, expectations_mbedtls_ssl_recv_t, int,
    (void * /* ctx */, unsigned char * /* buf */, size_t /* len */))
SETUP_POST_CHECKS((ssl_recv_fn))}
\end{lstlisting}%
\end{lrbox}
\begin{figure}[t]
  \scalebox{1.0}{\usebox{\figmbedtlsbox}}
\caption{\code{ssl_recv_fn} mock used in the unit proof for \code{ssl_msg_fetch_input}.}
\label{fig:fn_ssl_recv}
\vspace{-0.3cm}
\end{figure}%

\begin{table}[t]
\scriptsize
\centering
\begin{tabular}{|l|c|c|c|c|}
\hline
\textbf{Project} & \textbf{Unit Proofs(U)} & \textbf{Weeks(W)} & \textbf{Persons(P)} & \textbf{$\textbf{Rate} =U/(W \times P)$} \\
\hline
\mbedtls (vMocks) & 31 & 5 & 1 & $6.2$ \\
\hline
aws-c-common & 171 & 24 & 3 & $2.4$ \\
\hline
firecracker-vmm & 27 & 30 & 2 & $0.5$ \\
\hline
\end{tabular}
\caption{Comparison of engineer efficiency across verification projects.}
\label{table:engineer_eff}
\vspace{-1cm}
\end{table}

\bparagraph{Outcome and learning.}
A single verification engineer with no previous familiarity with the codebase was able to verify \verifiedfuncs functions is the SSL message component of \mbedtls in five weeks.
To put this into perspective, we looked at similar projects using the unit proof methodology of code-level verification for \bmc but does not use a DSL based mocking methodology -- the aws-c-common verification project from AWS~\cite{DBLP:journals/spe/ChongCEKKMSTTT21} in C and the Firecracker Virtual Machine Monitor verification project~\cite{KaniFirecracker} in \rust, again from AWS.
The comparative rate of developing unit proofs is given in~\cref{table:engineer_eff}.
The properties under verification and the complexity of the codebase vary across projects.
Thus it is hard to make an apples-to-apples comparison.
These numbers are only indicative that vMocks may be useful, especially bootstrapping a verification project.

With \mbedtls, it was challenging to determine pre-and-post conditions for functions
since most of the considered functions did not specify them in comments or employ unit tests.
In essence, verifying legacy code using unit proof faces similar problems as faced by developers retrofitting unit tests into such codebases.
The next section discusses the use case for vMocks based on industrial experience with tMocks.

\section{Usability of Mocks}
\label{sec:use}


\bparagraph{tMocks and TDD.}
tMocks are an essential tool for test driven development (TDD).
TDD is a methodology where unit tests are written before the \sut function.
The test initially fails since the function is not implemented.
The function and tests are then refined in lockstep until all the unit tests pass.
tMocks play a vital role in this refinement process since they specify how the function interacts with the environment.
There has been continued discussion on the efficacy of TDD over the last twenty years~\cite{beck2003test,DBLP:conf/ease/MunirWPM14,DBLP:conf/esem/RomanoZBPS22}.
TDD is not a silver bullet to improve software quality.
However, used judiciously, it has shown to improve software quality outcomes~\cite{DBLP:journals/ese/NagappanMBW08}.
Moreover, it is a promoted methodology of code development in industry~\cite{WrightSWEGoogleBook}.

In TDD, tMocks inform the design of object interfaces.
As~\cite{DBLP:conf/oopsla/FreemanMPW04a} states,
``By testing an object  in  isolation,  the  programmer  is  forced  to  consider  an  object’s interactions with its collaborators in the abstract, possibly before  those  collaborators  exist.  TDD  with  Mock  Objects  guides  interface  design  by  the  services  that  an  object  requires,  not  just  those  it  provides.''
Thus, tMocks compel the \sut to evolve into a form that is robust and well tested.
In contrast, current approaches to formal verification assume the \sut has already reached a robust design, having undergone sufficient testing.
It is in this context that formal verification is applied to software, often by a team of verification specialists who are different from the original developers~\cite{DBLP:conf/osdi/CadarDE08,DBLP:journals/spe/ChongCEKKMSTTT21,DBLP:conf/nfm/HamzaFKNS22}.

Using the lessons of TDD, this work present vMocks, to lower the startup costs of formal verification such that developers themselves use formal verification early on in software projects.
With mocking, the program under verification and the environment need not be fully fleshed out to gain confidence in design and implementation.
\sut functions can be verified in isolation as they are developed, i.e., using verification driven development (VDD).

\bparagraph{Lifecycle of Mocks.} We know from industrial experience that tMocks become difficult to maintain once the project reaches maturity and environment interfaces are fully fleshed out\footnote{\url{https://abseil.io/resources/swe-book/html/ch13.html\#prefer\_realism\_over\_isolation}}.
At this point, investing time and effort in fakes is better for the long term.
We imagine that vMocks will have a similar lifecycle.
They will be invaluable early on in a project when developers are
designing a system incrementally or refactoring a legacy codebase.
However, once a \sut is fully fleshed out with the aid of mocking,
using fake environments will be natural and serve the project better in the long run.
We see VDD supplementing TDD since the startup cost of TDD will be lower.
Where VDD will find its utility relative to TDD is unexplored.
This paper contributes a methodology and tooling to conduct research in this area.

\afterpage{\clearpage} 
\section{Related Work}
\label{sec:related}
This work stands on both research in formal verification and the practice of test driven development. 
Designing abstract environments is an integral part of any code-level formal verification effort.
Environments were described using function summaries in the static driver verification project (SDV)~\cite{DBLP:conf/eurosys/BallBCLLMORU06} from Microsoft Research.
As another design example, a fake filesystem  was implemented by the \klee authors for verifying coreutils~\cite{DBLP:conf/osdi/CadarDE08}.
Angelic verification~\cite{DBLP:conf/fmcad/LahiriLGNLKDLB20} from MSR, is a methodology to auto generate environments.
It was a response to the long lead times in delivering software environments in the SDV project.
However, automatic generation of environments is a program synthesis problem and is inherently hard for complex environments. 

In test driven development, the need to test with an environment before it can be defined led naturally to tMocks~\cite{DBLP:conf/oopsla/FreemanMPW04a}. Their applicability with respect to test driven development has been studied over the last two decades~\cite{beck2003test,DBLP:conf/ease/MunirWPM14,DBLP:conf/esem/RomanoZBPS22}.
More interestingly, tMocks have been widely promoted in industry. There is a large body of advice of how to use them, including caveats~\cite{WrightSWEGoogleBook}.
In this vein, the reader may find various ToTT~\cite{ToTTAnnounce} blog posts (e.g.,~\cite{ToTTPartialMocks,ToTTMockWhatYouOwn}) from Google interesting.
It is hard to maintain parity between a tMock and the real environment as a project evolves.
To mitigate the effort required to maintain tMocks, there has been interest in academia~\cite{DBLP:conf/cav/BraggFRS21} and industry (e.g.,~\cite{SourceryUrl,JestUrl}) on automatically generating tMocks.
Similar techniques may be useful in generating vMocks.

Software verification, including modeling environments for verification, remains a specialized endeavour as highlighted in a recent case study by AWS~\cite{DBLP:journals/spe/ChongCEKKMSTTT21}.
To our knowledge, there has been no published work on adapting testing mocks (tMocks) for verification (vMocks). 
This presentation takes the first step towards remedying the situation.

\section{Conclusion}
\label{sec:conclusion}
This work takes inspiration from mocks in testing (tMocks) and introduces their counterpart in verification (vMocks).
Like tMocks, vMocks define environments behaviorally.
We hope that vMocks catalyze formal methods in software development. 
The presented case study uses vMocks for writing proofs using vMocks for public functions of the SSL messaging layer of the \mbedtls project with low engineering effort.
For the case study, we developed \seamock, an open-source vMock DSL library embedded in C++ for the \seahorn \bmc tool.
A future direction would be to adapt static analysis of tools like \seahorn to reason about runtime polymorphism efficiently enabling the use of tMock frameworks like gMock in symbolic execution.
Alternatively, the use of \seamock could be extended to concrete testing.

\afterpage{\clearpage} 
\bibliography{bib}
\ifarxivmode
	\newpage
	\appendix
	\section{Appendix}
\label{app:vmock-appendix}

\catcode`\_=12 
\begin{table}[h]
\centering
\begin{filecontents*}{assets/key_to_testname_flipped.csv}
Key,Test name
1,ssl_msg_flight_free
2,ssl_msg_fetch_input
3,ssl_msg_recv_flight_completed
4,ssl_msg_handle_message_type
5,ssl_msg_transform_free
6,ssl_msg_dtls_replay_update
7,ssl_msg_set_outbound_transform
8,ssl_msg_get_record_expansion
9,ssl_msg_read
10,ssl_msg_flight_transmit
11,ssl_msg_start_handshake_msg
12,ssl_msg_set_inbound_transform
13,ssl_msg_parse_change_cipher_spec
14,ssl_msg_prepare_handshake_record
15,ssl_msg_decrypt_buf
16,ssl_msg_write_record
17,ssl_msg_dtls_replay_check
18,ssl_msg_buffering_free
19,ssl_msg_write_change_cipher_spec
20,ssl_msg_write
21,ssl_msg_close_notify
22,ssl_msg_flush_output
23,ssl_msg_encrypt_buf
24,ssl_msg_send_alert_message
25,ssl_msg_write_handshake_msg_ext
26,ssl_msg_check_record
27,ssl_msg_pend_fatal_alert
28,ssl_msg_finish_handshake_msg
29,ssl_msg_send_flight_completed
30,ssl_msg_read_record
31,ssl_msg_handle_pending_alert
\end{filecontents*}
\csvautotabular{assets/key_to_testname_flipped.csv}
\caption{Key to testname map for~\cref{fig:mbedtls_perf}.}
\label{tab:key_to_testname}
\end{table}

The \notinterestedfuncs function that either were specifications themselves or did not have memory accesses were~\code{mbedtls_ssl_reset_in_out_pointers}, \code{mbedtls_ssl_update_in_pointers},
\code{mbedtls_ssl_update_out_pointers}, \code{mbedtls_ssl_write_version}.

\section{Background}
\label{app:background}

\newsavebox{\figturtle}
\begin{lrbox}{\figturtle}%
	\begin{lstlisting}[language=C++, escapechar=@, numbers=left]
class Turtle {
  ...
  virtual ~Turtle() = default;
  void PenUp() {...}
  virtual void PenDown() = 0;
};
\end{lstlisting}%
\end{lrbox}%

\newsavebox{\figmockturtle}
\begin{lrbox}{\figmockturtle}
	\begin{lstlisting}[language=C++, escapechar=@, numbers=left]
#include "turtle.h"       // Brings in Turtle class
#include "gmock/gmock.h"  // Brings in gMock.
#include "gtest/gtest.h"

class MockTurtle : public Turtle {@\label{line:gmock_turtle_mock_class}@
 public:
  ...
  MOCK_METHOD(void, PenDown, (), (override));
};
using ::testing::AtLeast;

TEST(PainterTest, CanDrawSomething) {
  MockTurtle turtle;
  EXPECT_CALL(turtle, PenDown()).Times(AtLeast(1));@\label{line:gmock_turtle_expect}@
  Painter painter(&turtle);
  EXPECT_TRUE(painter.DrawCircle(0, 0, 10));
}
 \end{lstlisting}
\end{lrbox}

\begin{figure}[t]
	\subcaptionbox{Class to be mocked.\label{fig:turtle_gmock_example}}[0.45\linewidth]{%
		{\scalebox{1.0}{\usebox{\figturtle}}}}
	\subcaptionbox{A gMock usage example.\label{fig:gmock_example}}[0.45\linewidth]{%
		{\scalebox{1.0}{\usebox{\figmockturtle}}}}
	\caption{An adaptation from GoogleTest documentation.}
	\vspace{-0.5cm}
\end{figure}
\bparagraph{GoogleMock DSL.}
It is useful to see how expectations are setup in a mocking DSL.
We thus illustrate with an example usage of GoogleMock which is a popular mocking framework for C/C++.
The DSL language is typical of other mocking frameworks also.
~\cref{fig:turtle_gmock_example} shows a \code{Turtle} class that is the environment we want to test against.
The \sut is the \code{Painter} class. The function to be unit tested is~\code{Painter::DrawCircle}, which needs the \code{Turtle} environment.
The unit test is shown in~\cref{fig:gmock_example}.
We create a mock \code{Turtle} at~\lref{line:gmock_turtle_mock_class}.
The methods to be mocked are defined using \code{MOCK_METHOD}.
Moving to the actual unit test \code{CanDrawSomething}, a behavior expectation on the mock turtle is defined at~\lref{line:gmock_turtle_expect}.
We expect that the subsequent \code{painter.DrawCircle} call will invoke the \code{Turtle::PenDown} method at least once.
If this expectation is not met then the test fails.

From the example, we make two key observations.
First, the tMock is partial, i.e., it need not define all methods of a class. For example, \code{PenUp} is not mocked since we never expect it to be called in the unit test.
Second, a tMock specifies only how the environment is to behave with external observers; it does not pretend to mimic the environment operationally.

Overall gMock works well in the context it was designed for since:
\begin{inparaenum}[(1)]
	\item the gMock DSL enables concise specification of environment behaviours in terms of actions,
	\item the DSL uses standard C++ tooling enabling a familiar coding environment for the same developers who write production code, and,
	\item the framework produces executable code that fits together with the unit proof to make a closed testing environment.
\end{inparaenum}

\bparagraph{Bounded Model Checking and \seahorn.}
Code level Bounded Model Checking (\bmc) is a path sensitive and precise technique to verify software implementations.
The \emph{bounded} part of the name indicates that the states of a program are checked only to a certain depth.
Thus, loops are bound to a depth determined by the user.
\bmc is used to check safety properties of software.
In case a bad state is entered (within bounds), the technique generates a finite length counterexample showing how the state was reached.

The \seahorn \bmc engine~\cite{DBLP:conf/fmcad/PriyaSBZVG22} is a bit precise bounded model checker for \llvm programs based on the above principles.
It uses \texttt{Clang} to compile C/C++ programs to \llvm bitcode.
It provides a number of builtins to enable writing specifications in the style of C/C++ functions, for example, \code{sassert} and \code{assume} codify verification assertions and assumptions respectively.
Undefined but declared functions starting with \code{nd_} return non--deterministic values of the declared return type.
The \code{is_deref} builtin checks if a memory access is within allocated bounds.
The \seahorn pipeline automatically adds an \code{is_deref} check before every memory access.
The pair of \code{reset_modified},~\code{is_modified} builtins mark and check unexpected mutation of buffers designated as read-only.
The \code{memhavoc} builtin \emph{havocs} a newly created memory allocation, filling it with non--deterministic byte(s).

\newsavebox{\figsutexbox}
\begin{lrbox}{\figsutexbox}%
	\begin{lstlisting}[style=SeaC, escapechar=@, numbers=left,xleftmargin=.2\textwidth]
 unsigned water(
      uint32_t qty) {
  uint32_t p = 0;
  while (p < qty) {
    p += pour();
  }
  return p;
}
\end{lstlisting}%
\end{lrbox}

\newsavebox{\figenvexbox}
\begin{lrbox}{\figenvexbox}%
	\begin{lstlisting}[style=SeaC, escapechar=@, numbers=left,xleftmargin=.2\textwidth]
  uint32_t pour() {
    uint32_t v =
      nd_uint32_t();@\label{line:env-create-nd}@
    assume(v < 3);
    return v;
  }
\end{lstlisting}%
\end{lrbox}

\newsavebox{\figproofexbox}
\begin{lrbox}{\figproofexbox}%
	\begin{lstlisting}[style=SeaC, escapechar=@, numbers=left,xleftmargin=.2\textwidth]
  int main(void) {
    //pre
    uint32_t e = nd_uint32_t();@\label{line:proof-create-nd}@
    assume(e <= 10);@\label{line:proof-assume-nd}@
    //call SUT
    uint32_t p = water(e);
    //post
    sassert(p >= e);@\label{line:proof-check}@
  }
\end{lstlisting}%
\end{lrbox}

\begin{figure}[t]
	\subcaptionbox{Water plant \sut.\label{fig:sut_example}}[0.3\linewidth]{%
		{\scalebox{1.0}{\usebox{\figsutexbox}}}}
	\subcaptionbox{Environment for \sut.\label{fig:env_example}}[0.3\linewidth]{%
		{\scalebox{1.0}{\usebox{\figenvexbox}}}}
	\subcaptionbox{Unit proof for \sut.\label{fig:proof_example}}[0.3\linewidth]{%
		{\scalebox{1.0}{\usebox{\figproofexbox}}}}
	\caption{An example unit proof setup in a plant watering program.}
	\label{fig:water-plant}
	\vspace{-0.5cm}
\end{figure}

\begin{figure}[t]
	\includegraphics[trim={0cm 16.5cm 0cm 5.5cm},clip,scale=0.60]{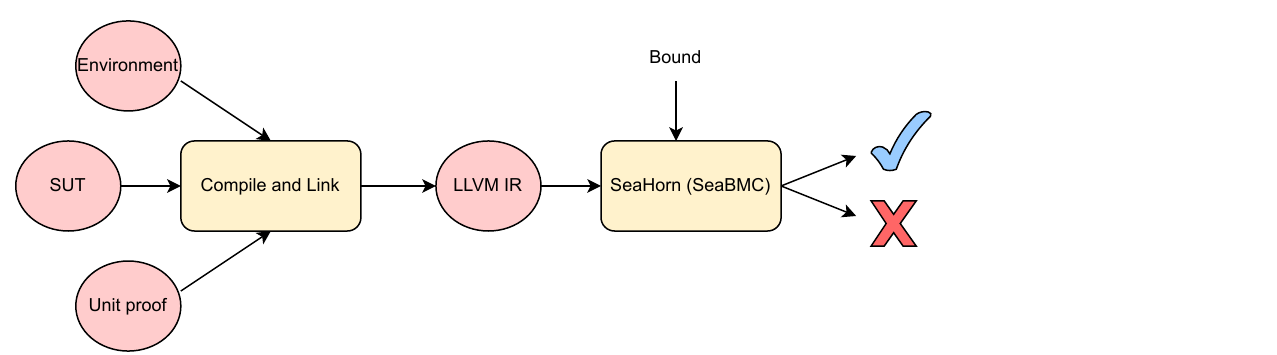}
	\caption{How a proof is assembled. \seahorn outputs pass (UNSAT) or fail (SAT). }
	\label{fig:seahorn-flow}
	\vspace{-0.7cm}
\end{figure}

\bparagraph{Unit proof.}
\bmc needs a harness to setup and check pre--and--post conditions respectively, and, call the \sut with valid arguments passed to the chosen entry point.
A methodology for designing harnesses on a per function basis in discussed in~\cite{DBLP:journals/spe/ChongCEKKMSTTT21}. In~\cite{DBLP:journals/isse/PriyaZSVBG22}, this harness specification is called a \emph{unit proof}, after unit tests.

We explain the concept with an example in C/C++ shown in~\cref{fig:water-plant}.
The \code{water} function in~\cref{fig:sut_example} is the \sut.
It takes in a quantity of water to pour on the plant and calls an environment function \code{pour} that releases small amounts of water to the plant repeatedly in a loop.
The loop exits when the required quantity of water has been released.
The \code{pour} environment function is defined in~\cref{fig:env_example}.
It creates a non--deterministic unsigned value in~\lref{line:env-create-nd}, constraints it using an \code{assume} and returns one of \code{0},~\code{1}, or, \code{2}.
Finally, the setup is closed by a unit proof in~\cref{fig:proof_example} inside a C/C++ \code{main} entrypoint function.
First the function sets the pre--condition that the quantity of water is between \code{0} and \code{10} in ~\lref{line:proof-create-nd} and~\lref{line:proof-assume-nd}.
It then calls \code{water} and checks the post--condition that atleast the required quanity of water was released. Note the following aspects of unit proofs.
\begin{enumerate}
	\item[(1)] The specification language for pre-and-post conditions is C/C++, the same language as the function under verification.
		This is for ease of developers already familiar with the target language.
		A detailed rationale is found in~\cite{DBLP:journals/isse/PriyaZSVBG22}, which calls it Code-As-Specification (CaS).
	\item[(2)] In this example \code{pour} is considered the environment.
		However, where to partition the \sut and the environment is a design choice.
		At one extreme, all immediate callee functions of an \sut may be an environment.
		Alternatively, an obvious lower layer (e.g., OS API) may be the environment.
		Though, in all cases, an environment must be chosen for the verification to be practical.
	\item[(3)] The precision of \bmc usually comes at a cost of scalability.
		As increasing amounts of code is brought into the purview of a unit proof, tool running times become impractical.
		Thus, constructing a unit proof requires engineering to get the right amount of code covered within the limits of tools.
\end{enumerate}
The third aspect is supported by data from two recent projects that use the unit proof methodology.
The verification of the aws-c-common library from AWS resulted in 171 unit proofs and was developed by 3 engineers over 24 weeks~\cite{DBLP:journals/spe/ChongCEKKMSTTT21}.
Another project from AWS that verified the firecracker virtual machine monitor produced 27 unit proofs in 30 weeks using 2 engineers~\cite{KaniFirecracker}.
These are long lead times to build confidence in software implementations where product release cadence can be a couple of weeks to a couple of months~\cite{DBLP:journals/tse/KulaGDG22}.
In this paper, we show that mocks can ease writing unit proofs and, hopefully, shorten development times for verification projects.

\Cref{{fig:seahorn-flow}} illustrates how the \sut, environment, and, unit proof is assembled for verification by \seahorn.
First the different parts are compiled and linked together using the \texttt{Clang} compiler toolchain to create a closed \llvm IR program.
Then, \seahorn is run with a bound set by the user.
For the \code{water} unit proof, the bound will be set to 10.
\seahorn generates verification conditions in the \smt language.
Finally, these are given to a solver (e.g., z3), which returns either UNSAT (proof holds) or SAT (proof has a counterexample).
In the case a counterexample is found, \seahorn generates a program trace of how the safety property is violated.

\else
\fi
\end{document}